\RequirePackage{snapshot}
\documentclass[
    reprint,
    superscriptaddress,
    amsmath,amssymb,
    aps,
    pra,
]{revtex4-2}
\renewcommand{\doi}[1]{\href{http://doi.org/#1}{doi:#1}}
\newif\ifshowlabels
\showlabelsfalse
\newif\iffigdraft
\figdraftfalse
\newif\ifgentoc
\newif\ifeditmode
\editmodefalse
\gentocfalse
\newif\ifshowparagraphinline
\showparagraphinlinefalse
\ifeditmode\else
\showparagraphinlinefalse
\fi
\usepackage[sectionbib]{bibunits}
\defaultbibliographystyle{new}
\defaultbibliography{references}
\usepackage{layouts}
\usepackage[acronyms]{glossaries}
\usepackage[procnames]{listings}
\usepackage{amsfonts}
\usepackage{blindtext}
\let\oldblindtext\blindtext
\renewcommand{\blindtext}{\textcolor{gray}{\oldblindtext}}
\usepackage{cmap}
\usepackage{datetime}
\usepackage{environ}
\usepackage{etoolbox}
\usepackage{booktabs}
\usepackage[utf8]{inputenc}
\DeclareUnicodeCharacter{03BD}{\ensuremath{\nu}}
\DeclareUnicodeCharacter{2212}{--}
\DeclareUnicodeCharacter{03BC}{\textmu}
\DeclareUnicodeCharacter{03B2}{\ensuremath{\beta}}
\DeclareUnicodeCharacter{03B1}{\ensuremath{\alpha}}
\DeclareUnicodeCharacter{03B3}{\ensuremath{\gamma}}
\DeclareUnicodeCharacter{00B0}{\ensuremath{^\circ}}
\DeclareUnicodeCharacter{00B1}{\ensuremath{\pm}}
\DeclareUnicodeCharacter{2192}{\ensuremath{\rightarrow}}
\DeclareUnicodeCharacter{03C3}{\ensuremath{\sigma}}
\DeclareUnicodeCharacter{03C4}{\ensuremath{\tau}}
\DeclareUnicodeCharacter{2009}{ }
\DeclareUnicodeCharacter{2080}{\ensuremath{_0}}
\DeclareUnicodeCharacter{2082}{\ensuremath{_2}}
\DeclareUnicodeCharacter{2084}{\ensuremath{_4}}
\DeclareUnicodeCharacter{1F449}{\ensuremath{\rightarrow}} 
\DeclareUnicodeCharacter{1F448}{\ensuremath{\leftarrow}} 
\iffigdraft
    \usepackage[draft]{graphicx}
\else
    \usepackage{graphicx}
\fi
\usepackage{hyphenat}
\usepackage{ifxetex}
\usepackage{mathrsfs}
\usepackage{calc}

\ifeditmode
\usepackage{pdfcomment}

\else
\newcommand{\pdfcomment}[2][]{}

\fi
\usepackage{hyperref}
\hypersetup{colorlinks=true, pdfstartview=FitV, linkcolor=linkcolor, citecolor=dbluecolor, urlcolor=dgreencolor, bookmarksdepth=subparagraph}
\usepackage{bookmark}
\usepackage{sidecap}
\usepackage{soul}
\usepackage{textcomp}
\usepackage{url}
\usepackage{xcolor}
\usepackage{xspace}
\usepackage{multirow}
\usepackage{array}
\newcolumntype{R}{>{\vspace{2ex}\displaystyle}{r}}
\newcolumntype{L}{>{\displaystyle}{l}}
\definecolor{SUgrey}{HTML}{6F777D}
\definecolor{SUorange}{HTML}{D44500}
\definecolor{dbluecolor}{rgb}{.01,.02,0.29}
\definecolor{dgraycolor}{rgb}{0.50,0.50,0.50}
\definecolor{dgreencolor}{rgb}{0.0,0.4,0}
\definecolor{linkcolor}{cmyk}{0,0.7,0.5,0.5}
\usepackage{accsupp}    
\newcommand{\noncopynumber}[1]{%
    \BeginAccSupp{method=escape,ActualText={}}%
    #1%
    \EndAccSupp{}%
}

\usepackage[compact]{titlesec}
\ifshowparagraphinline
\titleformat{\paragraph}[runin]{\color{gray}\normalfont\bfseries\footnotesize}{}{3pt}{\hspace{0.75em}\ul{\footnotesize\thesubsection\theparagraph)\;}}[:]
\else
\renewcommand{\paragraph}[1]{\par\phantomsection\addcontentsline{toc}{paragraph}{#1}}
\fi
\makeatletter
\@ifundefined{linelabel}{%
\newcommand{\linelabel}[1]{}}{}
\makeatother
\makeglossaries
\renewcommand{\glossarysection}[2][]{}
\include{myacronyms}
\renewcommand{\thesection}{\Roman{section}}
\renewcommand{\thesubsection}{\thesection.\arabic{subsection}}

\makeatletter
\renewcommand{\p@subsection}{}
\renewcommand{\p@subsubsection}{}
\makeatother
\newcounter{subfigure}[figure]
\newcounter{subfigurenonumber}
\newcounter{tempfigure}
\renewcommand\thesubfigurenonumber{(\alph{subfigurenonumber})}
\newcommand{\subfig}[2]{%
    \setcounter{tempfigure}{\value{figure}}%
    \addtocounter{tempfigure}{1}%
    \refstepcounter{subfigure}%
    \setcounter{subfigurenonumber}{\value{subfigure}}%
    \expandafter\edef\csname ref#2\endcsname{\thesubfigurenonumber}
    \label{#1}%
    }
\newif\ifpoormancref
\poormancreffalse
\ifpoormancref
\usepackage[poorman]{cleveref}
\else
\usepackage{cleveref}
\fi
\crefname{equation}{Eq.}{Eqs.}
\crefname{table}{Table}{Tables}
\crefname{figure}{Fig.}{Figs.}
\crefname{section}{Sec.}{Sec.}
\crefname{subfigure}{Fig.}{Figs.}
\crefname{lstlisting}{listing}{listings}
\Crefname{lstlisting}{Listing}{Listings}
\usepackage{xr}
\ifshowlabels
\usepackage{refcheck}
\makeatletter
\newcommand{\refcheckize}[1]{%
  \expandafter\let\csname @@\string#1\endcsname#1%
  \expandafter\DeclareRobustCommand\csname relax\string#1\endcsname[1]{%
    \csname @@\string#1\endcsname{##1}\wrtusdrf{##1}}%
  \expandafter\let\expandafter#1\csname relax\string#1\endcsname
}
\def\@setmarginlbl{%
    \if@show@ref
        \if@labelled
            \set@fbox@par
            \if@unsdlbl
                \makebox[0pt][l]{\zero@height{$\,$\rotatebox{90}{\scalebox{0.6}{\mark@size
                {\bfseries\upshape?}\underline{\last@lbl}{k\bfseries\upshape?}}}}}%
            \else
                \makebox[0pt][l]{\zero@height{$\,$\rotatebox{90}{\scalebox{0.6}{\fbox{{\mark@size\last@lbl}}}}}}%
            \fi
        \else
            \if@show@unl@bld
                \makebox[0pt][l]{\zero@height{$\,$\rotatebox{90}{\scalebox{0.6}{\unl@bld@mark}}}}%
            \fi\fi
        \fi
        \global\@labelledfalse
    }
\def\@setnmmarginlbl{%
    \if@show@ref
        \set@fbox@par
        \if@unsdlbl
            \hbox to \textwidth{\makebox[0pt][r]{\rotatebox{90}{\scalebox{0.6}{\mark@size{\bfseries
                            \upshape?}$\langle$\last@lbl$\rangle${\bfseries
            \upshape?}}}$\,$}\hfill}%
        \else
            \hbox to \textwidth{\makebox[0pt][r]{\rotatebox{90}{\scalebox{0.6}{\mark@size$\langle$%
            \last@lbl$\rangle$}}$\,$}\hfill}%
        \fi
    \fi
    \global\@labelledfalse
}
\def\@bibitem@proceed@#1{%
    \@ifundefined{cit@#1}{\@warning@rc@{Unused bibitem `#1'}%
        \if@show@cite
            \gdef\@biblabel{\makebox[0pt][r]{\zero@height{\rotatebox{90}{\scalebox{0.6}{{\mark@size{\bfseries\upshape?}}%
                \underline{\@verbatim@{#1}}{\mark@size{\bfseries\upshape?}}}}$\,$}}%
            \@@@biblabel@@}%
        \fi
    }{%
        \if@show@cite
            \set@fbox@par
            \gdef\@biblabel{\makebox[0pt][r]{\zero@height{\rotatebox{90}{\scalebox{0.6}{\fbox{TESTTESTTEST\@verbatim@{#1}}}}$\,$}}\@@@biblabel@@}%
        \fi
}}%
\makeatother
\refcheckize{\cref}
\refcheckize{\Cref}
\fi
\begin{document}
\newlength\myfigwidth
\setlength{\myfigwidth}{3.5in}
\newcommand\SUaffil{\affiliation{Department of Chemistry, Syracuse University, Syracuse, NY 13210, USA}}
\author{Farhana Syed}
\author{Jessica N. Khuc}
\author{Alexandria Guinness}
\author{John M. Franck}
\SUaffil
\email{jmfranck@syr.edu}
\title{Contiguous Patches of Translational Hydration Dynamics on the Surface of K-Ras}
\date{Preliminary arxiv draft: \today}

\begin{abstract}
Proteins involved in signaling pathways represent an interesting target
for experimental analysis by \gls{odnp} (Overhauser Dynamic Nuclear
Polarization), which determines the translational mobility at the
surface of proteins. They also represent a challenge, since the
hydration dynamics at all sites remains relatively rapid, requiring
sensitive measurements capable of drawing finer distinctions. Targeting
the protein K-Ras, we find \gls{odnp} cross-relaxivity values that appear
consistent within similar regions of 3D space, regardless of the
specific residue where the spin probe used to select the location has
been attached. The similar dynamics observed from nearby residues
indicate a persistence/uniformity of the translational dynamics of water
on the nanometer scale. This results makes sense, since it essentially
means that the dynamics of water remains consistent over a lengthscale
(a nanometer) over which liquid water exhibits structural persistence
(\emph{i.e.} its correlation length). This opens up the possibility of
strategically and comprehensively mapping out the hydration layer in
aqueous solution and identifying regions that contribute significantly
to the free energy of binding interactions -- for example, slow water
that might contribute significant entropy, or regions with strongly
temperature-dependent water mobility that might contribute significant
enthalpy.

\end{abstract}

\begin{bibunit}
\maketitle
\glsresetall
\section{Introduction}

This publication presents the experimental characterization of a
significant fraction of the surface hydration water that coats the K-Ras
signaling protein. Hydration water, present inside and on the surface of
proteins, displays significantly different characteristics from bulk
water, and these differences in properties likely contribute both to the
thermodynamics that drive interactions, as well as to kinetic barriers
\cite{Barnes2017, Chong2017, FranckMethEnz2018, Laage2017, Schiro2015}.
Ultimately, by better characterizing this water one can better
understand -- and therefore control -- the reactions and interactions of
biological macromolecules.

In this context, it is important to distinguish between different
categories of macromolecule-associated water molecules. ``Bound water''
remains hydrogen-bonded (or otherwise trapped) in position even in
crystal or powder form \cite{Ball2008}. Previous research has directly
demonstrated how some key binding interactions displace such structural
bound water molecules \cite{Lam1994RatDesPot, Young2007}. Internal
water resides in the cavities of most globular proteins, and evolution
conserves some of these much as it does the amino acids
\cite{Halle2004ProHydDyna}.

Although structural/bound/buried water molecules are clearly of import,
the layer of hydration water that envelops protein surfaces, typically
2-3 molecules deep and known as surface or interfacial hydration water
(or simply the ``hydration layer''), also demonstrates fascinating
differences from bulk water \cite{ContiNibali2014_JACS}. Furthermore,
despite the fact that water molecules in the interfacial layer
constantly exchange positions with bulk water on a relatively fast
timescale, they nonetheless exhibit a distinct pattern of varying
properties. This unique structure, or ``fingerprint,'' remains
consistent within the protein's reference frame, even amid the constant
flux of water molecules from hydration layer to bulk and \emph{vice
versa} \cite{FranckMethEnz2018, Dahanayake2019}. The hydration layer
affects the structure, dynamics, and function of the protein, and is
involved as a mediator of protein-protein, and protein-ligand
interactions \cite{Ball2008}.

Interfacial water is a vital and active constituent of biomolecules, at
least to the extent that it helps to determine protein structure
\cite{Tanford1962}, while some would go as far as to claim that it
actively drives a host of motions
\cite{Fenimore2002, Frauenfelder2009}. As with structural water,
displacement of interfacial water has been shown to be an important
factor for stronger ligand binding. Among the different protein surface
characteristics, solvation/desolvation characteristics stand out as
strong predictors of protein binding interfaces
\cite{Burgoyne2006PreProInt, Fiorucci2010PreProInt}. When the surface
area of a protein is buried by interaction with another protein, the
solvent accessible surface is lost and this loss (desolvation) has been
shown to be a driving force for strong protein-ligand binding. Partial
desolvation is required for binding as well as providing stability of
macromolecular intermediate complexes \cite{Camacho1999FreEneLan}.
Compelling evidence from simulations demonstrated that the
hydrogen-bonding rearrangements of the interfacial waters are
extraordinarily slow at protein surfaces that engage in strong
protein-protein binding interactions \cite{Chong2017}. Desolvation
forms one of two high energy barriers in the \gls{md} simulations of drugs
targeting G-protein coupled receptors (GPCRs) \cite{Dror2011PatMecDru}.
Besides the expected barrier that is due to the receptor geometry that
the ligand traverses before it reaches the binding pocket, an unexpected
earlier energy barrier was found located at the surface of the receptor
which was suggested to be due to the process of dehydration of both
ligand and receptor. The dehydration leads to greater hydrophobic
contacts which ascertains the entry of the ligand into the vestibule of
the receptor thus preventing its escape back into the bulk water. It has
been proposed that during the drug-design process water should be
included at the interface as it can mediate hydrogen bonds that lead to
higher specificity and affinity \cite{Ladbury1996}. A rational drug
design method followed on this proposition where the ligands were
structurally modified to enable it to form hydrogen bonds with the
interfacial water which lead to enhancement of scores that predict
ligand binding affinity \cite{Huang2008ExpOrdWat}.

We hypothesize that Ras proteins represent an extreme case with respect
to the importance of the hydration layer in guiding and determining
interactions. Paradoxically, on the one hand, researchers have
traditionally struggled to find small molecule binding sites on the
surface of Ras proteins
\cite{Cox2014_Ras, Spiegel2014, Moore2020RASTheUnd}, while on the other
hand, these proteins bind extensively to other proteins as part of their
role in signaling pathways \cite{Vetter2001, Moore2020RASTheUnd}. But,
the specific role of the interfacial hydration layer of K-Ras in binding
interactions remains unclear: Does desolvation actively contribute
entropy to the free energy to offset other costs? Or is the interfacial
water layer of Ras more bulk-like, simply offering a penalty-free
situation in which ligands can form fruitful hydrogen bonds to the K-Ras
surface?

Since the hydration water constitutes the first site of contact between
the protein and its binding-partner or a ligand, it plays a role in
relaying information about the protein to the incoming interactant. By
serving as a carrier of protein information, hydration water must
therefore exhibit different kinetic and thermodynamic properties at
different locations along the protein surface. Additionally, to
establish a stable binding, the two interactants must undergo
displacement of hydration water. Therefore, understanding the hydration
map of the protein will provide us insight into not only the structure
and function of a macromolecule but more importantly, the thermodynamic
properties of the hydration water will reveal the thermodynamics of
binding of ligands and interactants. Slow moving water molecules likely
serve as an entropic reservoir, and can be identified by their slow
rotational correlation times \cite{Denisov_DNA}. However, translational
mobility/diffusivity is an experimentally measurable property that
proves particularly intriguing for understanding important motions of
proteins. The translational motion likely correlates with changes to
larger-scale motions, such as conformational changes and binding
interactions \cite{Schiro2015}; it potentially reflects on the energy
required to slow down the diffusing water and break the hydrogen bonds
at the interface in order to bind to an incoming ligand or interactant.

Overhauser Dynamic Nuclear Polarization (\gls{odnp}) offers the unique
capability of isolating the contribution of translational dynamics in
the presence of bulk water, and with site-specificity. It is a
dual-resonance technique that uses nuclear magnetic resonance (\gls{nmr}) to
study the motion of hydrogen nuclei of water while also employing the
sensitivity of electron spin resonance (\gls{esr}) to enhance the \gls{nmr} signal
in the presence of bulk water. \gls{odnp} provides information on relaxation
processes that are measured as 2 types of relaxivities: \(k_{low}\) and
\(k_\sigma\) \(k_{low}\) lays out slower hydration water dynamics near
the magnetic field of the electron present on the spin label which is on
10-100s of nanosecond time-scale whereas \(k_\sigma\) gives us
information on faster hydration dynamics that are on 10-100s picosecond
or faster. \(k_{low}\) is modulated by the waters that are bound and
rotating with the protein while ksigma reflects on the hydration waters
that move around in the hydration layer due to diffusion or translation.
\cite{FranckMethEnz2018}

Therefore, we have employed \gls{odnp} to study the hydration dynamics of the
N-terminal portion of a globular protein K-Ras. The guanine
nucleotide-binding KRas proteins belong to a class of proteins known as
small GTPase that function as binary switches in cellular signal
transduction pathways that include regulation of cell growth,
differentiation, and proliferation \cite{Bourne1990GTPSupCon}. Ras
cycles between active guanosine-triphosphate (GTP)-bound and inactive
guanosine-diphosphate (GDP)-bound states \cite{Bos2007GEFGAPCri}.
GDP-bound form exists as a stable form in the resting cell. Upon
growth-factor related stimulation, it is converted to GTP-form with the
help of guanine nucleotide exchange-factors (GEFs) (Bos et al 2007). The
activated GTP-bound form transmits the signal to the downstream effector
proteins such as Raf, Ral-GDS, and PI3K, via direct binding to their
Ras-binding domains \cite{Simanshu2017RASProThe}. These effector
proteins carry out the sequence of events necessary for the purpose such
as cell growth or proliferation. Inactivation of the GTP-form is
achieved by GTPase-activating proteins (GAPs) which assist in the
hydrolysis of GTP to GDP (\cite{Scheffzek1998GTPProHel} Bos et al
2007). Mutated RAS remains constitutively activated in the GTP-Ras form,
losing control of regulation of signal transduction and thereby
resulting in uncontrolled growth
\cite{Bos1989RasOncHum, Wittinghofer2000RasMolSwi, Prior2012ComSurRas}).
About 20-30\% of all human cancers are caused due to mutations in
\emph{RAS} genes \cite{Bos1989RasOncHum, Prior2012ComSurRas}. K-Ras is
the most frequently mutated gene in most cancers, and K-Ras-4B is found
to be the most dominant isoform that is mutated in approximately 90\%
pancreatic cancers \cite{Biankin2012PanCanGen}, 30\% to 40\% of colon
cancers \cite{Neumann2009FreTypKRA}, and 15\% to 20\% lung cancers
\cite{Liu2019TarUntKRA}. We therefore employed K-Ras-4B, eliminating
the hypervariable region for simplicity, and built a nanometer-specific
map of the dynamics of the interfacial water coating K-Ras by generating
single cysteine mutants at 13 locations by site-directed mutagenesis.

\section{Experimental}

\subsection{Quantitative \gls{esr}}

Following accepted procedure \cite{E500Manual}, a \(\ge 20\;\mbox{μL}\)
in a 0.8~i.d. capillary tube (\(\sim 42\;\mbox{mm}\)) exceeds the volume
of the \gls{esr} cavity with any amount of sensitivity. As the data acquired
during \gls{esr} spectroscopy is a derivative of the absorption spectrum,
integrating twice (the double integral) yields a quantity proportional
to the number of spins in the sample. By comparing to a control
measurement of a known quantity of nitroxide label in aqueous solution
(typically hydroxy-tempo), one can quantify the concentration of
solution containing an unknown amount of spin label. Quantitative \gls{esr}
depends on accurate knowledge of the \(Q\)-factor of the resonator. As
we have found the XEPR determination of the Q factor to be imprecise and
unreliable, we employ a code snippet based on customxepr
\cite{customxepr} to capture the tuning curve (reflection profile) of
the resonator, and ensure that it matches that of the hydroxytempo
control.

\subsection{\gls{odnp}}

\gls{odnp} was performed following standard protocols.

\subsection{K-Ras expression and purification}

K-Ras plasmid was secured from addgene, and expression and purification
followed existing protocols. The expression was specifically optimized
to target an \gls{odnp} and \gls{esr} culture size, which is smaller than that of
\emph{e.g.} \gls{nmr} analysis and allows for parallel in-house (\emph{i.e.}
without a biochemical core facility) expression of up to 4 mutants at a
time.

\section{Results}

\subsection{\gls{mtsl} covalently binds to Ras at targeted sites}

K-Ras contains an N-terminal G domain that consists of 1-169 residues
and includes several conserved functional features known as switch
regions \cite{Pantsar2020CurUndKRA}. The C-terminal region known as
hypervariable region (HVR) helps anchor Ras to the cell membrane by
undergoing post-translational modifications at a cysteine. For
simplicity, we expressed only the G-domain, without the hypervariable
region, and substituted the 3 remaining native cysteines following
existing protocols \cite{Wind1999Ras} before expressing targeted
cyteine mutants that could accept precise placement of a stable
nitroxide radical-based spin-label
\emph{S}-(1-oxyl-2,2,5,5-tetramethyl-2,5-dihydro-1\emph{H}-pyrrol-
3-yl)methylmethanesulfonothioate (\gls{mtsl}).

In particular, the binding site of a macromolecular drug cyclic peptide
(cyclorasin 9A5) was mapped by
\textsuperscript{1}H-\textsuperscript{15}N Heteronuclear Single Quantum
Correlation (HSQC) \gls{nmr} spectroscopy \cite{Upadhyaya2015} and found to
be an extended area between switch I and switch II loops, within the
Ras-Raf interaction site, same site as small molecule ligands such as
DCAI \cite{Maurer2012SmaLigBin} that was identified by a fragment-based
screening \cite{Sun2012DisSmaMol, Maurer2012SmaLigBin}. The locations
for \gls{sdsl} (site-directed spin-labeling) were chosen near the cyclic
peptide binding site.

\begin{figure}
\hypertarget{fig:multMutStack}{%
\centering
\includegraphics[height=0.6\textheight]{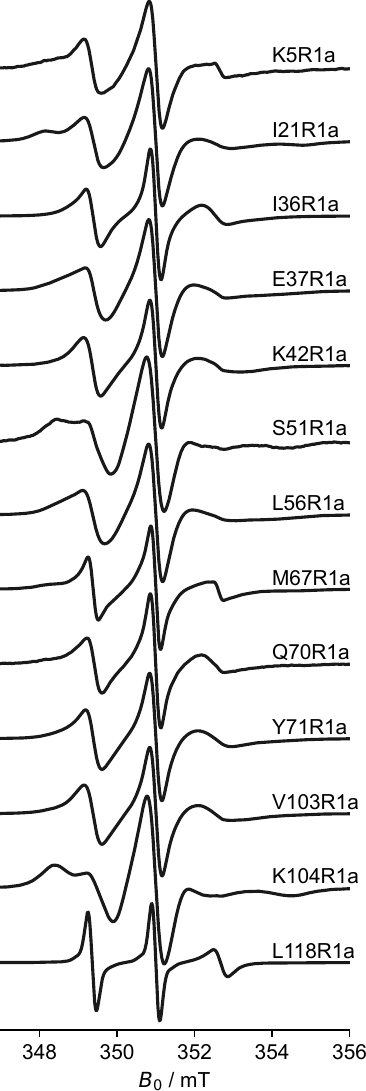}
\caption{R1a (\gls{mtsl} bound to a Cys) residue located at different sites
all generate \gls{esr} spectra that confirm complete attachment to the protein
by differing significantly from the three equal-amplitude peaks that
arise from free spin label. (As typical, this plot illustrates the
derivative of the \gls{esr} absorption with respect to magnetic field.) The
spectra all differ significantly from each other due to differences in
local side-chain packing that lead to different local dynamics of the
R1a sidechain. (Spectral amplitudes are normalized).
}\label{fig:multMutStack}
}
\end{figure}

\gls{sdsl} identifies the sites of interest on the protein's surface, forming
the basis for implementing \gls{odnp}. We first verify the success of \gls{sdsl} by
inspecting the \gls{esr} spectra of all the mutants. While \gls{esr} spectra of free
(not covalently attached) spin label typically yield three peaks whose
derivatives have equal height, spin labels covalently attached to
proteins yield spectra of distinctly different shapes.

It is worth noting that L118R1a in fig.~\ref{fig:multMutStack}
represents a spin label with qualitatively more dynamic behavior of the
spin label \emph{vs.} the other mutants; this may indicate that the
mutagenesis and spin labeling has disrupted the dynamics at this site,
but may also simply be reporting on significant flexibility at the
labeling site. Regardless, even in this case, the spectral data differs
significantly from the three equal derivative peaks expected from a free
spin label, clearly indicating covalent attachment of the \gls{mtsl} to the
cysteine.

N-terminal portion of K-Ras, known as G-domain, that contains 169
residues. K-Ras serves as a model for a fairly large sized soluble
protein that also has a significance as an oncogene. N-terminus of K-Ras
contains 3 native cysteines that were mutated to either serine or
leucine as these substitutions do not render the protein inactive
\cite{Sondermann2005}. On the cysteine-less K-Ras background 12 single
cysteine mutations were generated to attach the spin-label, \gls{mtsl}, in
order to study hydration dynamics around a single site at a time. The
residues measured are shown in fig.~\ref{fig:ksigmaPymol} in red and
blue.

A specific region of K-Ras participates in its function, specifically,
the residues located in the conserved switch regions are involved in
binding to the regulators GEFs and GAPs as well as its effector proteins
that are involved in the downstream signaling. The same region is
involved in drug binding. Small molecule drug ligand DCAI (Maurer et
al., 2012) and a macromolecular drug cyclic peptide, cyclorasin 9A5,
\cite{Upadhyaya2015} have been shown to bind to an extended area
between switches 1 and II. Cyclic peptide contact sites are shown in
Fig. 2. The residues selected for cysteine mutations for this study are
located in the cyclic peptide binding region.

CW X-band \gls{esr} spectrum of purified and spin-labeled K-Ras mutants were
collected to determine if K-Ras is spin-labeled with \gls{mtsl} in which case
it exhibits a broad line shape. Fig. 3 shows \gls{esr} spectra of K-Ras
mutants. The broad line shape of these spectra compared to the free spin
label confirms the spin-labeling of K-Ras mutants. Mutants exhibit
differences in the \gls{esr} spectra that can be broadly divided into 2
classes based on the presence or absence of the immobile component. The
size of immobile component can be used as the basis for further
distinguishing them into 3 categories: small immobile component, big
immobile component, and broad peak where the immobile component does not
show a clear distinct bump but continues to rise into the mobile
component. The presence of mobile and immobile component suggests a
distribution of rotamers that fall into two groupings with two different
order parameters (\emph{e.g.}, a spin label that spends more time at one
particular site, but occasionally escapes on wider trajectories).

The differences in \gls{esr} spectra relate to the orientation of the amino
acid residue (Fig. 2); if the residue faces outward then it displays
more mobility whereas a residue facing inward into the core of the
protein experiences reduction in its mobility giving rise to an immobile
component on the \gls{esr}.

\subsection{\gls{esr} provides concentration to enable relaxivity
calculations}

As noted in the theory section, \gls{odnp} in the presence of a fully
saturated \gls{esr} transition yields a quantity proportional to
\(C_{SL}k_\sigma\), where \(C_{SL}\) is the concentration of the spin
label while the cross-relaxation rate (\(k_\sigma\)) is the primary
quantity of interest for understanding the molecular dynamics of water.
Therefore, normalization by the concentration of the spin-label forms an
essential element of the experimental strategy here. As noted in the
Experimental section, the double integral of the \gls{esr} data provides the
concentration of spin label. However, we observe that 5-10\% variation
in the concentration reported by \gls{qesr} for the same sample is not
uncommon. relatively insignificant variations in the baseline of the \gls{esr}
spectra can lead to even for samples This is even true with the controls
mentioned in the experimental section and with customized code that
allows detailed control over the process of baseline correction and
double integration. Inspecting controlled cases of such variation
reveals very similar spectra with nearly identical amplitudes, and
reveals that relatively insignificant variations in the baseline of the
\gls{esr} spectra lead to the 5-10\% variation reported by \gls{qesr}.

As part of this study, we repeat the acquisition of \gls{esr} and \gls{odnp} data
for each \gls{sdsl} mutant several times, in order to average out any small
experimental discrepancies or small discrepancies in sample preparation.
This repetition also affords the opportunity to more accurately quantify
the \gls{qesr} concentration.

\begin{figure}
\hypertarget{fig:singleMutantEsr}{%
\centering
\includegraphics[width=\linewidth]{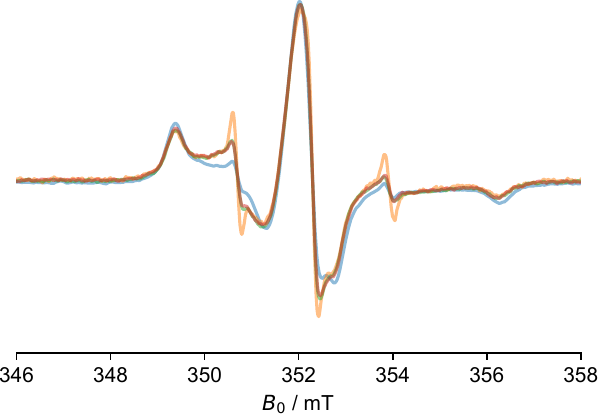}
\caption{In contrast to the difference observed between mutants in
fig.~\ref{fig:multMutStack}, independently expressed and purified
batches of K-Ras with the R1a sidechain attached at the same location
yield highly reproducible \gls{esr} spectra. A least-squares alignment and
normalization of spectra arising from 4 independent S175R1a samples
highlights subtle differences between samples. For example, the orange
spectrum displays a slight small spin label impurity that remained after
desalting (here corresponding to an exceedingly small percentage of the
total spin label in the sample). As another example, the blue spectrum
arises from a sample with a significantly higher concentration than the
others (in the mM range); here, differing ratio of mobile component
(350.7~mT) to immobile component (349.4~mT) indicates that
protein-protein interactions are significant enough to affect the
sidechain dynamics at this site (indicating unsuitability for \gls{odnp}
analysis).}\label{fig:singleMutantEsr}
}
\end{figure}

We find that many such issues can be rectified with a simple program
that determines the relative scaling of the different repeated datasets
needed to minimize the least-squares difference between the spectra as
shown in fig.~\ref{fig:singleMutantEsr}. First, this presentation
notably highlights small differences in the spectra -- whether arising
from small free spin label contaminant (acceptable) or protein-protein
interactions at higher concentration (not acceptable). Returning to the
problem of accurate spin label quantification, the average of the double
integral for all valid samples averages discrepancies due to small
baseline differences, while the scalings needed to determine the rms
difference provide the precise relative concentrations of the different
samples (avoiding issues with small baseline variations).

\subsection{Surface-Exposed Sites Yield Significant Hyperpolarization}

Even very raw data exhibits noticeable differences from transmembrane
protein samples or lipid bilayer samples that have been studied by \gls{odnp}
previously. Noticeably, even though the spin-labeled protein
concentrations in play are hundreds of micromolar, the samples here
routinely yield a signal enhancement of the water signal ranging between
\(-9\times\) to \(-25\times\) the Boltzmann polarization that an aqueous
sample yields in the absence of \gls{odnp}. Not surprisingly, the resulting
cross-relaxation values \(k_\sigma\), tend to be a significant fraction
of the 95~s\textsuperscript{-1}M\textsuperscript{-1}, the relaxivity of
a small molecule spin label dissolved in bulk water \cite{FranckPNMRS}.

\begin{table}
    \centering
    \begin{tabular}{cc}
        \hline
        Mutant & \(k_\sigma\) / s\textsuperscript{-1}M\textsuperscript{-1}\\
        \hline
        \hline
        I21R1a & 49 \\
        I36R1a & 73 \\
        E37R1a & 65 \\
        K42R1a & 56 \\
        S51R1a & 38 \\
        L56R1a & 70 \\
        M67R1a & 41 \\
        Q70R1a & 41 \\
        Y71R1a & 49 \\
        V103R1a & 73 \\
        L118R1a & 46 \\
        \hline
    \end{tabular}
    \caption{\label{tbl:ksigmaVals}Cross-relaxivity (\(k_\sigma\)) observed
    when different residues of K-Ras are substituted with R1a (followup
measurements underway). The ``R1a'' sidechain corresponds to a cysteine
with a covalently attached \gls{mtsl}.}
    \label{tab:ksigmaVals}
\end{table}

\gls{nmr} signal enhancement depends on the concentration of the spin-labeled
sample. Such enhancements show that we have a significant concentration
of spin-labeled sample that is suitable to obtain \gls{odnp} data.

\subsection{Cross-Relaxivity Measurements Demonstrate Consistency in 3D
Space}

\begin{figure}
\hypertarget{fig:ksigmaPymol}{%
\centering
\includegraphics[width=0.5\linewidth]{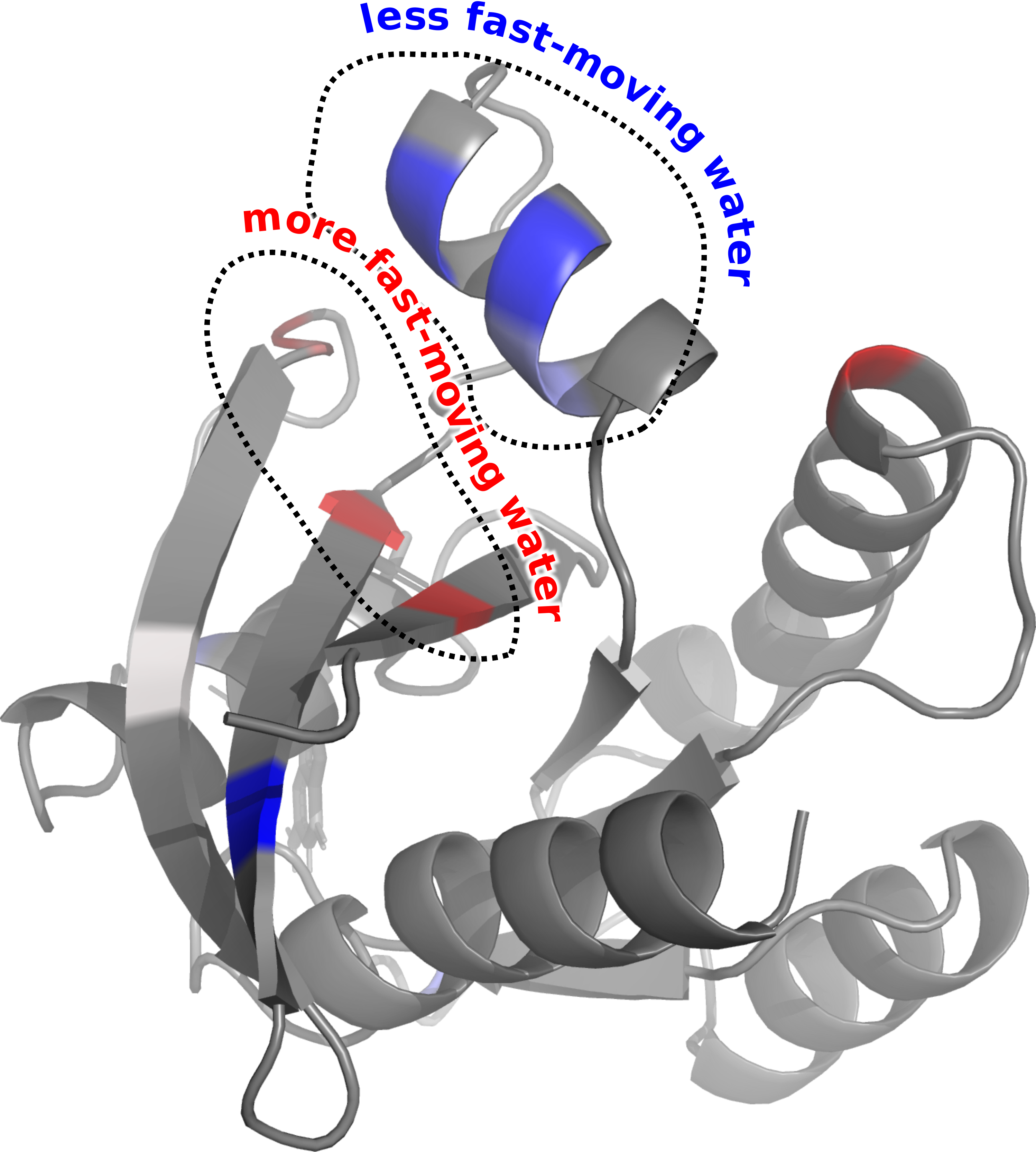}
\caption{Preliminary measurements of K-Ras G domain indicate contiguous
regions with fast \emph{vs} slow hydration dynamics. The red
\(\rightarrow\) blue color scheme indicates measured high
\(\rightarrow\) low \gls{odnp} \(k_\sigma\) values, from
tbl.~\ref{tbl:ksigmaVals}. Unmeasured residues are colored in grey. The
fact that we observe similar dynamics at nearby residues indicates a
persistence/uniformity of the 10-100~ps water dynamics on the nanometer
scale. By combining such nanometer patches, one can now comprehensively
map out the hydration layer in aqueous solution and identify regions
that contribute significantly to the free energy of binding interactions
-- for example, slow water that might contribute significant entropy, or
regions with strongly temperature-dependent water mobility that might
contribute significant enthalpy.}\label{fig:ksigmaPymol}
}
\end{figure}

Beginning this study, we were interested in observing the dynamics of
interfacial hydration water near the binding sites of the cyclic
polypeptide cyclorasin 9A5. As the location of the R1a sidechains
determines which portion of the hydration water is observed with each
measurement, the location of the R1a sidechains were chosen primarily as
outward-facing sidechains (\emph{i.e.} mutants where the native
amino-acid sidechain facing the solvent) near the binding site of
cyclorasin 9A5 \cite{Upadhyaya2015}. The chosen residues served as
probes looking into the water dynamics near the region that interact
with the cyclic peptide. As a result, the R1a sidechains of the
resulting samples were both surface-exposed and clustered in a similar
region of space.

For example, one might assume that the interfacial hydration water in
this region moved particularly slowly, and that the desolvation of the
surface offered up an entropic contribution to aid in the binding of the
cyclic polypeptide. However, as noted in tbl.~\ref{tbl:ksigmaVals}, much
of the cross-relaxation contributions represent a significant fraction
of the 95~s\textsuperscript{-1}M\textsuperscript{-1} cross-relaxation
rate of a small spin label in water. The cross-relaxation rate,
\(k_\sigma\), is specifically sensitive to the fast motion on 100s of ps
timescale. It depends on the motion of the water molecules that diffuse
fast near the spin label within 5-10 Å sphere around the nitroxide spin
label. If the residue is exposed to the water, we expected to see a
faster diffusion of water near the spin and therefore a higher value of
\(k_\sigma\), whereas a buried residue would give yield a lower
\(k_\sigma\). Therefore, the present result means that there are nearly
as many water molecules moving nearly as fast as in bulk. We do not see
the 5-fold slow-downs observed \emph{e.g.} with previous studies of CheY
and certainly not the more dramatic slowdowns seen with Annexin XII
\cite{Barnes2017, FranckMethEnz2018}. With this data in hand, we are
reminded that many Ras binding events are predicted to be primarily
enthalpically driven, so that a very large contribution to the entropy
of binding from desolvation might not be required or expected. In other
words, this might represent an intriguing experimental case where --
perhaps because of the relative smoothness of its surface -- K-Ras does
not offer the opportunity to release free energy in the form of entropy
through desolvation of its surface.

As noted in the introduction, a key motivation behind these measurements
is to understand how desolvation (removal of the interfacial hydration
water to make way for binding partners or ligands) contributes to the
free energies of binding of binding partners or lead molecules. As
already noted, the measurements support the conclusion that in the case
of K-Ras, this desolvation doesn't encourage binding through very large
entropic contributions. That said, they nonetheless point out that some
limited contribution could

More broadly, however, fig.~\ref{fig:ksigmaPymol} indicates how spin
labels exposed to the surface at nearby sites observe similar dynamics.
Patches of residues adjacent to each other in space report \(k_\sigma\)
values that are similar, and indicate the persistence of similar
dynamics on the nanometer scale. This results makes sense, since it
essentially means that the dynamics of water remains consistent over a
lengthscale (a nanometer) over which liquid water exhibits structural
persistence (\emph{i.e.} its correlation length). This encourages a
strategy where dense labeling, as in the present case or with even
denser or more extensive \gls{sdsl}, can map out these continuous patches of
water, while future measurements can then interrogate how the different
discrete patches respond to -- \emph{e.g.} changes in temperature,
binding or ligands, or deliberate manipulation of the hydration
properties through limited engineering of surface-displayed side-chains.

\section{Conclusions}

While several transmembrane and membrane associated proteins have been
studied with \gls{odnp}, relatively few studies have focused on the dynamics
of the hydration water coating the surface of globular signaling
proteins. Because of its importance, and the long-standing challenges it
has posed to drug design, we were motivated to scrutinize K-Ras through
a new lens, focusing on the behavior of hydration water on its surface.
In particular, we were motivated to perform measurements on several
mutants that were designed to display R1a (\emph{i.e.} spin label)
sidechains on the surface where various drug candidates (lead compounds)
or binding partners are known to bind.

The fact that we observe similar dynamics at nearby residues on the
surface of K-Ras indicates a persistence/uniformity of the 10-100~ps
water dynamics on the nanometer scale. By strategically choosing just
one or two labeling sites in order to observe a particular patch, and
then combining results from such nanometer patches, one can now
comprehensively map out the hydration layer in aqueous solution and
identify regions that contribute significantly to the free energy of
binding interactions -- for example, slow water that might contribute
significant entropy, or regions with strongly temperature-dependent
water mobility that might contribute significant enthalpy.

\appendix
\putbib
\end{bibunit}
\newif\ifsuppinfo
\suppinfofalse
\ifsuppinfo
\pagebreak
\onecolumngrid
\begin{center}
\makeatletter
\textbf{\large Supplemental Materials for:\\
\@title}
\makeatother
\end{center}
\twocolumngrid
\begin{bibunit}
\setcounter{equation}{0}
\setcounter{figure}{0}
\setcounter{table}{0}
\setcounter{page}{1}
\makeatletter
\renewcommand{\theequation}{S\arabic{equation}}
\renewcommand{\thefigure}{S\arabic{figure}}
\renewcommand{\bibnumfmt}[1]{[S#1]}
\renewcommand{\citenumfont}[1]{S#1}
\glsresetall
\input{supp_content}
\putbib
\end{bibunit}
\fi
\end{document}